\documentclass[a4paper,10pt,twocolumn]{article}
\usepackage[utf8]{inputenc}
\usepackage{multicol,graphicx}
\usepackage{mathptmx}
\usepackage[T1]{fontenc}
\usepackage{textcomp}
\usepackage{url}
\usepackage{amssymb}

\usepackage[T1]{fontenc}
\usepackage[english]{babel}

\usepackage[backend=bibtex,style=ieee,doi=false,isbn=false,url=true,maxnames=6,citestyle=numeric-comp,giveninits=true]{biblatex}

\fontfamily{ptm}\selectfont

\bibliography{limited_refs.bib}
\usepackage[font=small]{caption}
\usepackage{amsmath}

\newcommand{\mysection}[1]{\vspace{0.4cm} \uppercase{#1} \vspace{0.4cm}}
\newcommand{\mysubsection}[1]{\hspace{10pt}\textit{#1:}}

\makeatletter

\makeatletter
\def\blfootnote{\gdef\@thefnmark{}% Delete the marker symbol entirely
  \bgroup\leftskip=0pt\noindent%   Force flush-left (no empty space indentation)
  \@footnotetext}
\makeatother

\begin{document}
	
\setlength{\textfloatsep}{10pt plus 1.0pt minus 2.0pt}	
\setlength{\columnsep}{1cm}

%----------------------------------------------------------------------------------------
% Header (Title, Authors, Facilities)
%----------------------------------------------------------------------------------------

\twocolumn[%
\begin{@twocolumnfalse}
\begin{center}
	{\fontsize{14}{18}\selectfont
		% Title
        \textbf{Deep Sleep Classification via EEG Signal Criticality: A Passive BCI Approach for Sleep-Improvement Neurofeedback
        }\\}
    \begin{large}
        \vspace{0.5cm}
        % Authors
        Stanisław Narębski\textsuperscript{1}, Tomasz Komendziński\textsuperscript{1}, Tomasz M. Rutkowski\textsuperscript{2,3,1} \\
        \vspace{0.2cm}
        % Facilities
        \textsuperscript{1}Nicolaus Copernicus University, Toruń, Poland\\
        \textsuperscript{2}Araya Inc., Tokyo, Japan\\
        \textsuperscript{3}The University of Tokyo, Tokyo, Japan \\
        \vspace{0.2cm}
        E-mail: stanislaw@bci-lab.info \& tomek@bci-lab.info
        \vspace{0.2cm}
    \end{large}
\end{center}	
\end{@twocolumnfalse}%
]%

\blfootnote{%\vspace{-1.0\baselineskip}
Accepted for publication in the Proceedings of the 10th Graz Brain-Computer Interface Conference 2026, Graz, Austria, September 14-17, 2026.}

%----------------------------------------------------------------------------------------
% Abstract
%----------------------------------------------------------------------------------------

ABSTRACT:
Automated sleep staging is a fundamental application of passive Brain-Computer Interfaces (pBCI), decoding spontaneous neural states to enable closed-loop interventions independent of user intent. This study evaluates criticality features derived from Detrended Fluctuation Analysis (DFA) for the specific identification of deep sleep (N3).

We analyzed $347,232$ EEG epochs from $290$ older women using UMAP manifold learning to visualize state transitions. Subsequently, six classifiers were benchmarked via 10-fold cross-validation, using balanced accuracy to determine the optimal "state-sensing" engine for neurofeedback.Naive Bayes achieved the highest mean balanced accuracy ($87.17\% \pm 0.24\%$), significantly outperforming a fully connected deep neural network (FNN: $81.58\%$) and Random Forest ($80.97\%$). Linear models (LDA: $57.21\%$; SVM: $51.01\%$) performed poorly, indicating that DFA-derived criticality features reside on a distinct, non-linear manifold.

Probabilistic decoding of EEG criticality provides a high-accuracy sensing mechanism for pBCIs. This robust classification pipeline supports the development of state-dependent neurofeedback, such as targeted auditory stimulation, to enhance cognitive recovery.

%----------------------------------------------------------------------------------------
% Introduction
%----------------------------------------------------------------------------------------

\mysection{introduction}

The conceptualization of Brain-Computer Interfaces (BCI) has traditionally been dominated by "active" paradigms~\cite{bciBOOKwolpaw,tomekFRONTIERS2016}, where users exert conscious effort to modulate neural activity for external control. However, as the field shifts toward everyday neurotechnology, the Passive BCI (pBCI) has emerged as a superior framework for state-monitoring and autonomous intervention~\cite{zander2011towards,tomekEMBC2019}. Perhaps the most robust, yet underappreciated, "textbook example" of a pBCI is automated sleep staging~\cite{apsipaBIOSIPS2015,tomekIIISsymposium2016}. Unlike active BCIs that require task-related intent, a sleep-based pBCI monitors the brain’s spontaneous, ongoing neural oscillations to infer physiological state without any voluntary participation from the user~\cite{stasCCN2025}.

By definition, a BCI is classified as "passive" when it satisfies three core criteria: the use of spontaneous brain activity as input, the absence of conscious intent from the user, and the output being a state estimation rather than a direct command~\cite{zander2011towards}. Sleep staging maps perfectly to this architecture. During sleep, the user is unconscious and unable to voluntarily modulate neural patterns. The system, therefore, must decode "hidden" physiological markers—such as sleep spindles, K-complexes, and slow-wave activity—to classify the internal state (e.g., N3 vs. REM)~\cite{sleepSTAGINGuptoN3_2017}.

In recent years, the utility of sleep staging has evolved from retrospective clinical diagnosis to real-time, "closed-loop" intervention~\cite{stasCCN2025}. This transformation is best exemplified by Closed-Loop Auditory Stimulation (CLAS)~\cite{CLASfoundation2013,CLASreview2016,CLASclinical2017}. In these systems, a pBCI architecture acquires EEG signals in real-time, identifies the specific "up-state" of a slow oscillation during deep sleep (N3)~\cite{sleepSTAGINGuptoN3_2017}, and triggers a millisecond-long auditory pulse. This intervention is designed to enhance the amplitude of slow waves, thereby improving memory consolidation and metabolic health~\cite{stasCCN2025}. In this context, the staging algorithm is the "decoder" of the BCI, providing the necessary state-inference to trigger a neurofeedback loop entirely below the level of conscious awareness~\cite{CLASfoundation2013,CLASreview2016,CLASclinical2017}.

The Critical Brain Hypothesis~\cite{beggs2008criticality} describes the idea that our nervous system organizes itself to be near a phase transition between order and disorder, a point analogous to the boundary between a solid and a liquid in physical systems. At this critical point, neural networks exhibit a characteristic set of emergent properties: optimal information processing, scale-free architecture (both in time as well as space), high sensitivity to external stimuli and flexibility in transitioning between states. Based on this we can predict that healthy brains will be closest to criticality with disease moving us away from this~\cite{massobrio2015criticality}. Furthermore, criticality has been shown to diminish during waking hours and to be restored during sleep~\cite{pearlmutter2009new}.

%Alzheimer/dementia criticality change
%Add limitation

While neural signals often exhibit complex heterogeneity, DFA provides a robust framework for quantifying long-range temporal correlations (LRTC) by identifying a single scaling exponent that governs signal fluctuations across time scales, which serve as a proxy for signal criticality. Previous studies indicate that disease states, specifically Alzheimer’s disease and related dementias, are associated with a breakdown in the brain's capacity to maintain flexible, long-range correlations, often resulting in a shift of the scaling exponent toward values indicative of decreased physiological complexity~\cite{tomekEMBC2022, 10.3389/fnhum.2023.1155194, tomekFAN2024}. By quantifying these shifts, we can evaluate the predictive utility of EEG criticality in identifying early-stage cognitive decline.

The current study builds upon the pBCI framework by leveraging EEG signal criticality~\cite{EEGandDFAfoundation_2001,tomekEMBC2021_2,tomekEMBC2022} as a primary feature set for deep sleep classification. Criticality—the study of neural systems operating at a phase transition between order and disorder—offers a unique window into the brain's self-organization during the descent into deep sleep~\cite{sleep2023criticality}. By utilizing criticality-derived metrics—specifically channel-wise DFA—as features for machine learning, we propose a high-accuracy, automated pipeline for the detection of N3 sleep stages. We further employ UMAP manifold learning to characterize the geometric structure of the resulting feature space, and benchmark six classifiers spanning linear, probabilistic, ensemble, and deep learning paradigms. This approach serves as the foundational "sensing" component for a subsequent sleep-improvement neurofeedback system, aiming to optimize the timing and efficacy of state-dependent interventions for the enhancement of human recovery and, ultimately, long-term cognitive health~\cite{stasCCN2025}. 
\begin{figure*}[h!]
	\centering
	\includegraphics[width=0.92\textwidth]{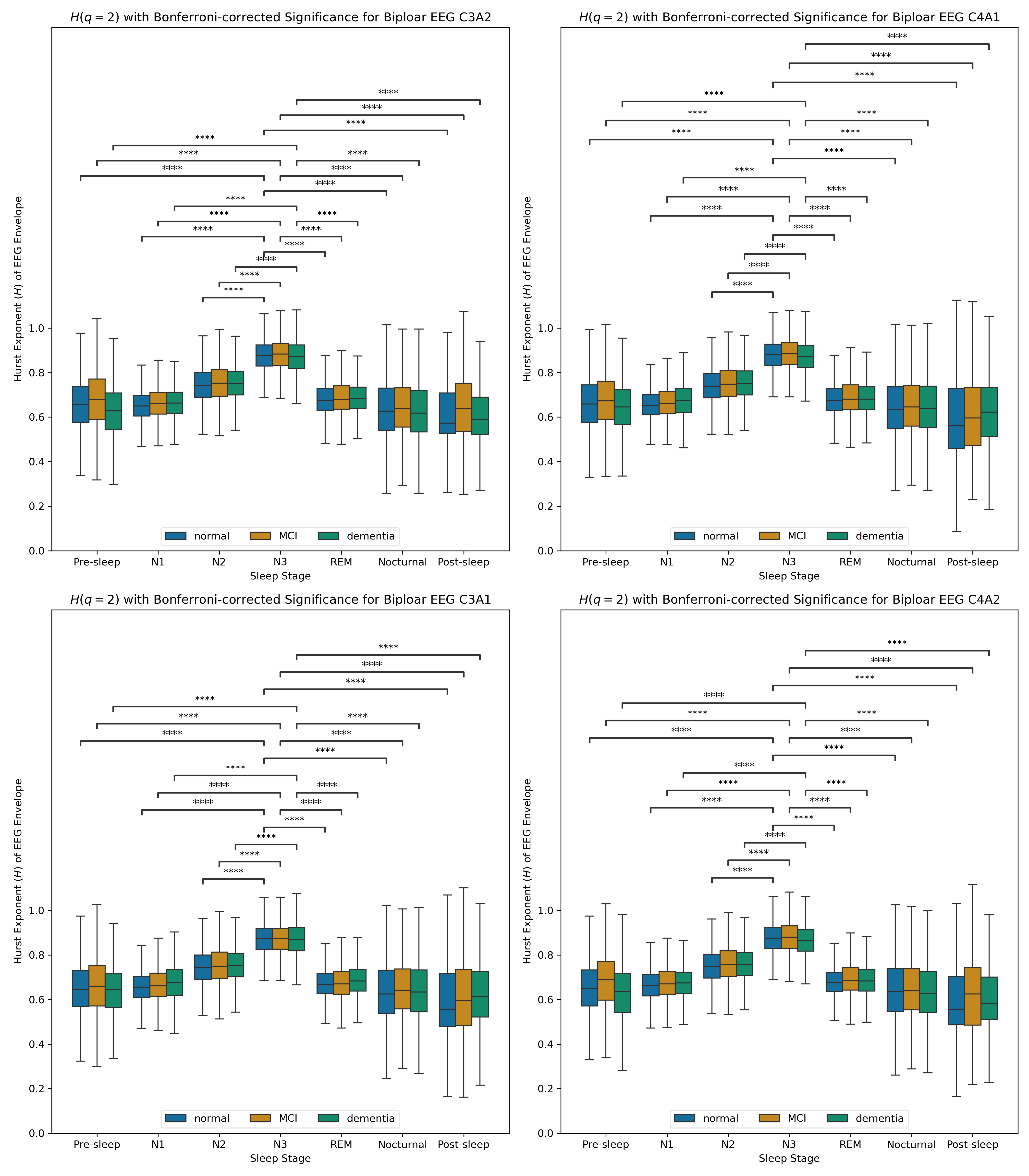}
	\caption{The Hurst exponent distributions ($q=2$) obtained from Detrended Fluctuation Analysis (DFA) of the EEG amplitude envelope for all sleep stages and cognitive groups. The four panels show values calculated from electrode pairs C3A2, C4A1, C3A1 and C4A2. Above the plots there are Bonferroni-corrected pairwise significance bars showing that the N3 stage exhibits significantly elevated $H$ values relative to all other sleep stages across all channels and cognitive groups (${*}{*}{*}{*}: p \leqslant 0.0001$).}
	\label{fig:dfa}
\end{figure*}

%----------------------------------------------------------------------------------------
% Materials and Methods
%----------------------------------------------------------------------------------------

\mysection{materials and methods}

The research dataset was acquired following proper authorization from the National Sleep Research Resource (NSRR)~\cite{zhang2018national}, drawing upon a selected portion~\cite{spira2008sleep} of the Study of Osteoporotic Fractures (SOF) records~\cite{cummings1990appendicular}. SOF participants were American women at least 65 old, with precise age unavailable due to de-identification. Overnight polysomnography (PSG) assessments were carried out in subjects' homes employing Compumedics P-Series equipment, which captured four electroencephalography (EEG) channels from 461 subjects. To establish a more uniform baseline and allow meaningful comparison, the present investigation included only those participants who obtained a Mini-Mental State Examination (MMSE)~\cite{folstein1975mini} score exceeding $24$ during PSG administration and who also attended a subsequent evaluation five years later ($N=290$). This cohort was divided into three groups according to their results on the Teng-modified Mini-Mental State Examination (3MS)~\cite{teng1987modified} administered at follow-up. Individuals obtaining a score of $88$ or above were designated as cognitively normal, those scoring between $78-88$ were identified as having Mild Cognitive Impairment (MCI), and those scoring below $78$ were classified as having dementia. The polysomnography electrode configuration comprised C$3$, C$4$, A$1$, and A$2$ channels (representing central and mastoid sites) with a sampling frequency of $128$~Hz. 

Subsequently, these recordings were divided into 30-second epochs and hand-scored by qualified technicians to classify wakefulness, rapid eye movement (REM) sleep, or non-REM phases (N1, N2, N3 and N4), following Rechtschaffen and Kales scoring standards~\cite{rechtschaffen1968manual}. For this study, N3 and N4 stages as originally scored in SOF were merged into a unified N3 stage consistent with contemporary clinical practice~\cite{iber2007aasm}. In addition, the awake state was partitioned into pre-sleep - starting from the first epoch until the first recorded sleep stage and post-sleep - the last awake state not followed by any more sleep. For six participants, a singular last sleep epoch was removed from analysis.

The following analysis received ethical approval from the Nicolaus Copernicus University in Toruń, Poland, Faculty of Philosophy Research Ethics Committee (Decision No. 60/2025).

\mysubsection{EEG Criticality Estimation with Detrended Fluctuation Analysis (DFA)}
%To characterize the non-stationary dynamics of sleep EEG, we employed Detrended Fluctuation Analysis (DFA)~\cite{DFA_2}. While neural signals often exhibit complex heterogeneity, DFA provides a robust framework for quantifying long-range temporal correlations (LRTC) by identifying a single scaling exponent that governs signal fluctuations across time scales. In the context of neurodegenerative pathology, the DFA scaling exponent serves as a proxy for signal criticality. Previous studies indicate that disease states, specifically Alzheimer’s disease and related dementias, are associated with a breakdown in the brain's capacity to maintain flexible, long-range correlations, often resulting in a shift of the scaling exponent toward values indicative of decreased physiological complexity~\cite{tomekEMBC2022, 10.3389/fnhum.2023.1155194, tomekFAN2024}. By quantifying these shifts, we aimed to evaluate the predictive utility of EEG criticality in identifying early-stage cognitive decline. 
DFA captures how signal fluctuations scale across time, yielding a single exponent that reflects the degree of long-range temporal correlations (LRTC) in EEG signal. In the context of sleep, this scaling exponent serves as a proxy for criticality: as the brain transitions into N3, theory predicts a characteristic shift toward long-range ordered dynamics, distinct from the activity during other sleep stages. We also aimed to verify whether the differences in criticality remain a reliable indicator of N3 regardless of cognitive status across healthy aging, early-stage cognitive decline and dementia.
The DFA procedure for each 30-second EEG epoch was implemented in four primary steps:
\begin{enumerate}
	\item \textbf{Integration:} The original time series $x(i)$ is transformed into an integrated profile $Y(i) = \sum_{k=1}^{i} [x(k) - \langle x \rangle]$ to convert the noise-like signal into a random walk-like profile, revealing underlying scaling properties.
	\item \textbf{Segmentation:} The integrated profile is partitioned into $N_n = \text{int}(N/n)$ non-overlapping segments of equal length $n$.
	\item \textbf{Local Detrending:} For each segment $\nu$, a local polynomial trend $y_{\nu,n}$ (of order $k=1$) is calculated via least-squares fitting and subtracted from the integrated profile to remove non-stationary "drifts."
	\item \textbf{Fluctuation Function Calculation (q=2):} The root-mean-square fluctuation is computed by averaging the variance over all segments:
	\begin{equation}
		F(n) = \sqrt{\frac{1}{2N_n} \sum_{\nu=1}^{2N_n} [F^2(\nu,n)]}.\label{eq:dfa}
	\end{equation}
\end{enumerate}
The output is characterized by the scaling exponent $\alpha$ (equivalent to the Hurst exponent $H$), determined as the slope of $\log F(n)$ versus $\log n$. A value of $\alpha \approx 0.5$ indicates white noise (randomness), while $\alpha \approx 1.0$ suggests $1/f$ noise, characteristic of healthy, self-organized criticality. Significant deviations from these values serve as primary metrics for pathological decline and loss of signal complexity.
\begin{figure}[t]
	\centering
	\includegraphics[width=\columnwidth]{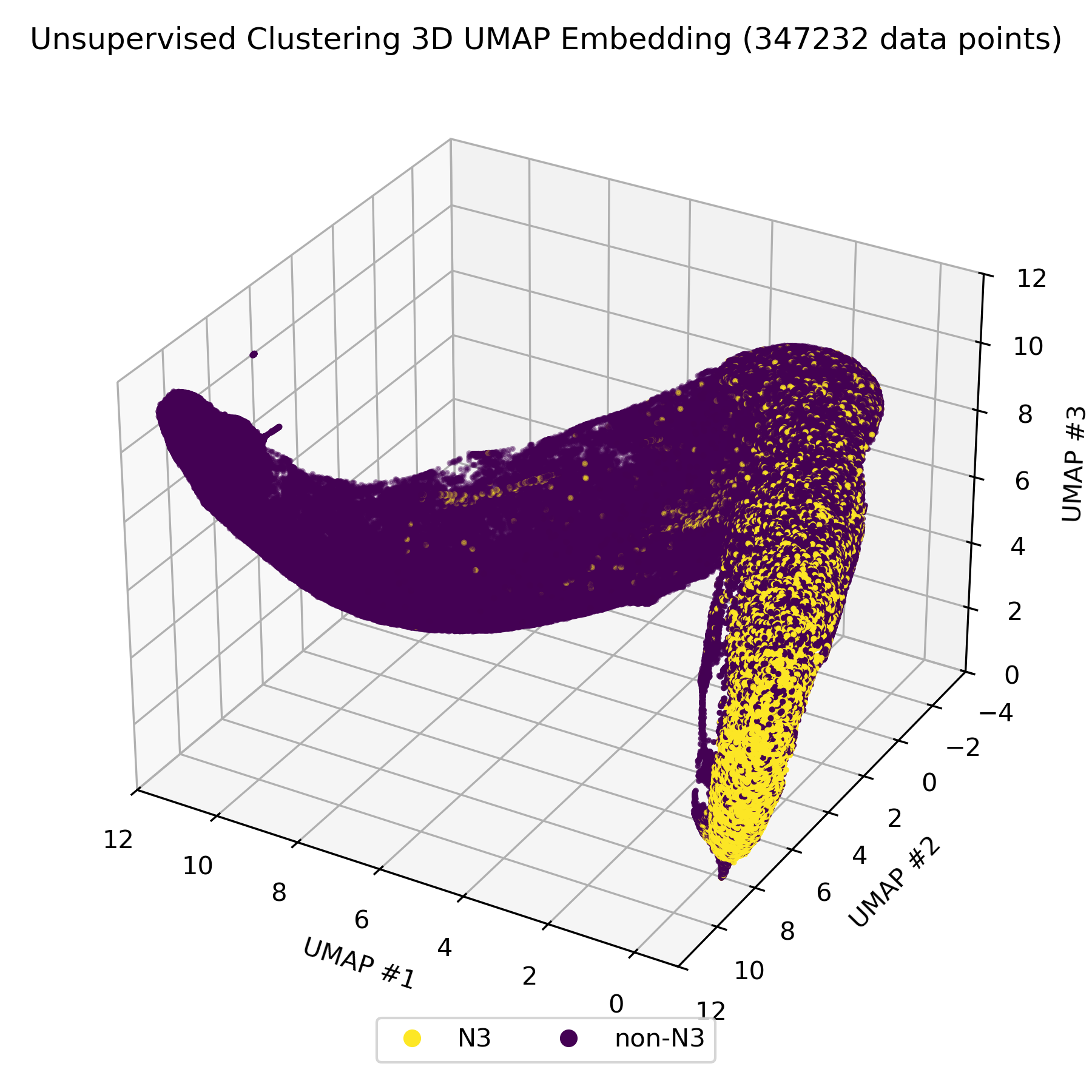}
	\caption{Unsupervised Uniform Manifold Approximation and Projection results for all 347,232 epochs, where yellow represents the N3 stage and purple non-N3 stages. The three-dimensional projection of the high-dimensional feature set created from DFA analysis of four-channel EEG sleep data reveals a distinct cluster of N3 epochs. The non-linear, curved topology of the embedding indicates that the N3 stage is not linearly separable from the rest of the sleep stages. }
	\label{fig:umap}
\end{figure}
\begin{figure}[t]
	\centering
	\includegraphics[width=\columnwidth]{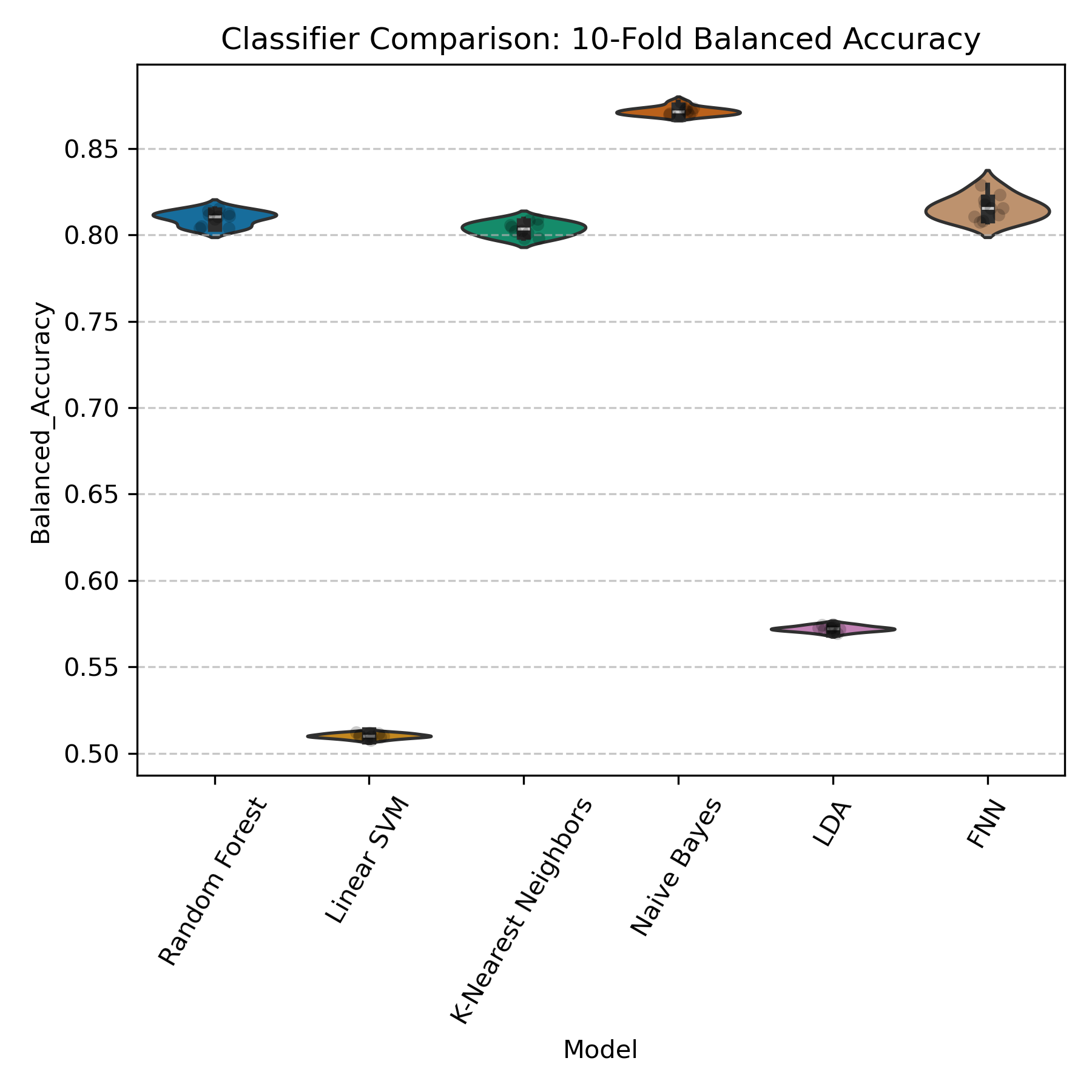}
	\caption{Balanced accuracy results from 10-fold cross-validation for binary classification of N3 versus non-N3 sleep across six evaluated classifiers: Random Forest, Linear SVM, K-Nearest Neighbors, Naive Bayes, Linear Discriminant Analysis (LDA), and a fully connected deep neural network (FNN). The highest score was achieved by Naive Bayes with accuracy of $87.17\% (\pm 0.24\%$). The full list of values can be seen in Table~\ref{tab:acc}.}
	\label{fig:acc}
\end{figure}

\mysubsection{Manifold Visualization of Sleep Dynamics}
To visualize the manifold structure of the DFA scaling features and assess their distribution across sleep architecture, we applied unsupervised Uniform Manifold Approximation and Projection (UMAP)~\cite{mcinnes2018umap}. Unlike supervised variants that rely on categorical labels, unsupervised UMAP constructs a low-dimensional embedding based solely on the underlying topological structure of the data. This allows for an unbiased observation of how neural scaling properties naturally cluster across different sleep stages. Data for this visualization were derived from the sleep polysomnography subset~\cite{spira2008sleep} of the Study of Osteoporotic Fractures (SOF) database~\cite{cummings1990appendicular, zhang2018national}. The resulting embedding highlights a clear topological separation of the N3 sleep stage from the continuum of awake, N1, N2, and REM sleep. Mathematically, UMAP optimizes a low-dimensional fuzzy simplicial set to match the high-dimensional topology. In our analysis, the N3 stage—characterized by high-amplitude delta activity and distinct long-range temporal correlations—occupies a restricted, high-density region of the manifold. This isolation reflects the unique "criticality" of deep sleep compared to the more stochastic or fragmented dynamics of lighter sleep stages. By examining these clusters, we can visualize how the breakdown of $q=2$ scaling exponents specifically impacts the N3 signature in the context of cognitive decline, where the once-distinct N3 cluster typically begins to merge with or shift toward less complex sleep states.

\mysubsection{Machine Learning Framework and Model Selection}
To evaluate the discriminative power of the DFA scaling exponents across different sleep stages, we implemented a comparative machine learning framework. We selected a diverse ensemble of six classifiers to ensure that our findings were robust to specific algorithmic biases, spanning linear, instance-based, probabilistic, and connectionist architectures:
\begin{description}
	\item[Linear Models:]  Linear Discriminant Analysis (LDA) and Linear Support Vector Classification (Linear SVC) were employed to establish a baseline for linear separability in the feature space. 
	\item[Non-Linear \& Ensemble Models:] Random Forest (100 estimators, max depth of 10) was used to capture hierarchical feature interactions, while K-Nearest Neighbors (KNN, $k=5$) assessed local topological clustering.
	% \item[Probabilistic \& Neural Models:] Gaussian Naive Bayes provided a baseline for conditional probability-based classification. Finally, a four-layer Feedforward Neural Network (FNN) with an architecture of (100, 100, 100, 50) neurons and ReLU activation was implemented to capture high-dimensional, non-linear dependencies.
	\item[Probabilistic \& Neural Models:] Gaussian Naive Bayes was employed to provide a baseline for conditional probability-based classification. Additionally, a four-layer Feedforward Neural Network (FNN) consisting of three 100-neuron hidden layers and a final 50-neuron layer with ReLU activation was implemented. This latter architecture was designed to capture high-dimensional, non-linear dependencies, following the successful application of a similar model in prior work~\cite{tomekEMBC2022}.
\end{description}
	
\mysubsection{Validation and Performance Metrics}
To mitigate the risk of overfitting and ensure the generalizability of the results, we employed a 10-fold Stratified Cross-Validation ($10 \times$ SKF) scheme. Stratification was used to preserve the class distributions (e.g., N3 versus non-N3 sleep) across each fold. Given the potential for class imbalance in sleep stage distribution and clinical diagnostic groups, model performance was quantified using Balanced Accuracy. This metric is defined as the arithmetic mean of class-specific recall:
	\begin{equation}
		\text{Balanced Accuracy} = \frac{1}{2} \left( \frac{TP}{TP + FN} + \frac{TN}{TN + FP} \right),\label{eq:bacc}
	\end{equation}
where $TP$, $FP$, $TN$, and $FN$ represent true positives, false positives, true negatives, and false negatives, respectively. All features were standardized using a StandardScaler within the cross-validation pipeline to ensure zero mean and unit variance, preventing models like SVC and FNN from being biased by differing feature scales.

%----------------------------------------------------------------------------------------
% Results
%----------------------------------------------------------------------------------------

\mysection{results}

Analysis of the Hurst exponent from DFA analysis of the EEG envelope ($H, q = 2$) across all $347,232$ epochs revealed a consistent and statistically robust elevation of H values during the N3 sleep stage relative to all other recorded states. As illustrated in Figure~\ref{fig:dfa}, this pattern was observed uniformly across all four bipolar EEG derivations (C3A2, C4A1, C3A1, and C4A2) and across all three cognitive groups (cognitively normal, MCI, and dementia). Bonferroni-corrected pairwise comparisons confirmed that N3 H values were significantly elevated compared to pre-sleep wakefulness, N1, N2, REM, nocturnal wakefulness, and post-sleep wakefulness (all $p \leqslant 0.0001$).

Unsupervised UMAP applied to the full 347,232-epoch DFA feature set produced a three-dimensional embedding that revealed a geometrically distinct cluster corresponding to N3 sleep epochs (Figure~\ref{fig:umap}). N3 epochs (shown in yellow) occupied a compact, high-density region sharply separated from the broader manifold of non-N3 sleep stages (shown in purple), which formed a more diffuse, curved continuum.

Six classifiers were evaluated on a binary N3 versus non-N3 classification task using ten-fold stratified cross-validation with balanced accuracy as the primary performance metric. Results are summarized in Table~\ref{tab:acc} and visualized in Figure~\ref{fig:acc}. Gaussian Naive Bayes achieved the highest mean balanced accuracy of $87.17\%~(\pm0.24\%)$, substantially outperforming all other evaluated models. The Feedforward Neural Network (FNN) attained the second-highest performance at $81.58\%~(\pm0.68\%)$, followed closely by Random Forest at $80.97\%~(\pm0.41\%)$ and K-Nearest Neighbors at $80.35\%~(\pm0.39\%)$. These four non-linear and probabilistic classifiers reached higher accuracy relative to the linear models. Linear Discriminant Analysis (LDA) achieved only $57.21\%~(\pm0.16\%)$ balanced accuracy, while Linear SVM performed the worst at $51.01\%~(\pm0.14\%).$ 
\begin{table}[t]
	\begin{small}
		\caption{Ten-fold cross-validation balanced accuracy results}\label{tab:acc}
		\begin{tabular*}{\columnwidth}{l @{\extracolsep{\fill}} ll}
			\hline
			Model                & Mean balanced accuracy  \\
			\hline                    
            Naive Bayes         & $0.871667 \pm 0.002397$ \\
            FNN                 & $0.815764 \pm 0.006825$ \\
            Random Forest       & $0.809687  \pm 0.004126$ \\
            K-Nearest Neighbors & $0.803512 \pm 0.003902$ \\
            LDA                 & $0.572146 \pm  0.001600$ \\
            Linear SVM          & $0.510112 \pm 0.001382$ \\
			\hline
		\end{tabular*}
	\end{small}
\end{table}

%----------------------------------------------------------------------------------------
% Discussion
%----------------------------------------------------------------------------------------

\mysection{discussion}

This study investigated the utility of EEG signal criticality as a feature set for automated deep sleep (N3) classification within a passive BCI (pBCI) framework. The findings demonstrate that criticality-derived features provide a high-accuracy, robust sensing mechanism for N3 detection, achieving a mean balanced accuracy of $87.17\%$ with Naive Bayes classification.
The elevated Hurst exponent values observed specifically during N3 sleep, consistently across all four EEG channels and all three cognitive groups, provide strong empirical support for the theoretical prediction that deep sleep serves a criticality-restorative function. Notably, this discriminative signature was preserved across cognitive status groups, suggesting that DFA-derived criticality remains a robust indicator of N3 sleep even in the presence of cognitive decline. 
While the exclusive inclusion of older women represents a demographic limitation that constrains the generalisability of these findings to the broader population, it simultaneously constitutes a strength in the context of the study's future aims - this cohort closely mirrors the target demographic for home-based sleep intervention and dementia prevention, where older women are both disproportionately represented and at elevated risk.
The UMAP embedding provided a critical insight that directly motivated and subsequently validated the classifier selection strategy. The geometrically distinct but topologically curved cluster occupied by N3 epochs in the three-dimensional embedding demonstrates unambiguously that deep sleep is not linearly separable from other stages in the DFA feature space. This curvature implies that the decision boundary required for accurate N3 detection is non-linear, which explains the failure of both linear models evaluated. The poor performance of the Linear SVM $(51.01\%)$ and the LDA $(57.21\%)$ models are indicative of a mismatch between model architecture and feature geometry.

%----------------------------------------------------------------------------------------
% Conclusion
%----------------------------------------------------------------------------------------

% \newpage
\mysection{conclusions}

The findings of this study reinforce the conceptualization of automated sleep staging as a premier use case for pBCI. By leveraging neural criticality—specifically through channel-wise DFA—we have demonstrated that spontaneous, unconscious oscillations contain high-fidelity signatures of physiological state. The robust performance of non-linear classifiers, such as the Naive Bayes model ($87.17\%$ balanced accuracy), validates that these criticality-derived features can reliably decode the "hidden" markers of N3 sleep across diverse cognitive profiles.
This architecture satisfies the definitive criteria of a pBCI: it utilizes spontaneous brain activity without conscious intent to produce a state estimation rather than a direct command. Our results show that the geometric complexity of the DFA feature space, as revealed by UMAP manifold learning, necessitates non-linear decoding strategies to accurately capture the brain's transition into deep sleep.
Ultimately, this pipeline provides the necessary ``sensing'' foundation for closed-loop interventions. By accurately identifying the N3 stage in real-time, such a system can trigger neurofeedback or auditory stimulation entirely below the level of conscious awareness. This study thus moves the field closer to a seamless, autonomous neurotechnology capable of monitoring and restoring neural criticality, thereby optimizing human recovery and long-term cognitive health.
%
%----------------------------------------------------------------------------------------
% References
%----------------------------------------------------------------------------------------

\mysection{references}
\printbibliography[heading=none]
\end{document}